\documentclass[letterpaper, 10 pt, conference]{ieeeconf}

\usepackage{graphicx}
\IEEEoverridecommandlockouts                 
\usepackage{amsfonts}
\usepackage{subfigure}
\usepackage{graphicx}
\graphicspath{{./IMG/}}
\usepackage{epstopdf}
\usepackage{cite}

\usepackage{color,array}
\usepackage{comment}
\usepackage{amsmath}
\usepackage{etoolbox}
\usepackage{algorithm,algorithmic}
\usepackage{mathrsfs}
\AtBeginEnvironment{algorithm2e}{\algoequations}
\AtEndEnvironment{algorithm2e}{\restoreequations}
\newcounter{algosavedequation}
\newcommand{\algoequations}{%
  \setcounter{algosavedequation}{\value{equation}+1}%
  \setcounter{equation}{0}%
  \renewcommand{\theequation}{\arabic{algosavedequation}\alph{equation}}
}
\newcommand{\restoreequations}{%
  \setcounter{equation}{\value{algosavedequation}}%
}

\allowdisplaybreaks[3]

\newtheorem{assumption}{Assumption}

\newtheorem{remark}{Remark}
\newtheorem{theorem}{Theorem}
\newtheorem{definition}{Definition}

\newtheorem{lemma}{Lemma}

\newcommand*{\QEDBL}{\hfill\ensuremath{\blacksquare}}
\newcommand\oprocendsymbol{\hbox{$\square$}}
\newcommand\oprocend{\relax\ifmmode\else\unskip\hfill%
\fi\oprocendsymbol}

\newcommand{\bx}{{\mathbf{x}}}

\newcommand{\by}{{\mathbf{y}}}
\newcommand{\ba}{{\mathbf{a}}}

\newcommand{\norm}[1]{\Vert #1 \Vert}

\begin{document}

\title{\bf Multi-Task System Identification of Similar Linear Time-Invariant Dynamical Systems}

\author{Yiting Chen, Ana M. Ospina, Fabio Pasqualetti, and Emiliano Dall'Anese
\thanks{Y. Chen, A. M. Ospina, and E. Dall'Anese are with the Department of
Electrical, Computer and Energy Engineering, University of Colorado Boulder; emails: \{yiting.chen-1, ana.ospina, emiliano.dallanese\}@colorado.edu. F. Pasqualetti is with the Department of Mechanical Engineering, University of California Riverside; email: fabiopas@engr.ucr.edu. }
\thanks{This work was supported in part by the National Science Foundation (NSF) through the awards 1941896 and 1926829, and by the Army Research Office (ARO) through the award W911NF1910360.}
}

\maketitle

\begin{abstract}%
This paper presents a system identification framework -- inspired by multi-task learning  -- to estimate the dynamics  of a given number of linear time-invariant (LTI) systems jointly by leveraging structural similarities across the systems. In particular, we consider LTI systems that model networked systems with similar connectivity, or LTI systems with small differences in their matrices. The system identification task involves the minimization of the least-squares (LS) fit for individual systems, augmented with a regularization function that enforces structural similarities. The proposed method is particularly suitable for cases when the recorded trajectories for one or more LTI systems are not sufficiently rich, leading to ill-conditioning of LS methods. We analyze the performance of the proposed method when the matrices of the LTI systems feature a common sparsity pattern (i.e., similar connectivity), and provide simulations based on real data for the estimation of the brain dynamics. We show that  the proposed method requires a significantly smaller number of fMRI scans to achieve similar error levels of the LS. 
\end{abstract}


\section{Introduction}
\label{sec:introduction}

System identification is a core task where the model of dynamical systems is estimated based on observed inputs and states~\cite{ljung1987theory,pillonetto2022regularized}. In particular,  identification  of linear time-invariant (LTI) systems is a well-investigated problem that has recently received renewed attention due to lines of research in the context of data-driven control and optimization (see, for example, the representative works in~\cite{de2019formulas,coulson2019data,hewing2020learning,berberich2020data,krishnan2021direct,li2022data}).

When the observation of the state is noise-free, the LTI system matrices can be estimated by leveraging the Willems’ Fundamental Lemma, provided that the recorded trajectory satisfies the persistency of excitation (PE) condition as discussed in, e.g.,~\cite{willems2005note,de2019persistency}. On the other hand, when process noise or disturbances enter the LTI system, several existing works focus on the asymptotic and finite time estimation errors  and sample complexity of the least squares (LS) estimator; see, for example, the representative works~\cite{sarkar2019near,simchowitz2018learning,faradonbeh2018finite,oymak2019non,zheng2020non,Xin2022learning} and pertinent references therein. 
Additionally, regularized system identification methods are investigated in, e.g.,~\cite{chen2014system,chiuso2019system,pillonetto2022regularized}; a low-order linear system identification via regularized regression is considered in~\cite{sun2020finite}. These regularized identification methods allow one to add a prior on the system matrices, and to strike a balance between LS fit and model complexity~\cite{hastie2009elements,hastie2015statistical}. 

The performance of the LS estimator hinges on the availability of a recorded trajectory that is sufficiently rich to render the LS well conditioned. In this paper, we  are interested in cases where the LS method is ill-conditioned. In particular, we consider the task of estimating the system matrices of $N > 1$ LTI systems, in cases where we do not have sufficiently long (and sufficiently rich) recorded trajectories for at least one of the systems (or for some of the systems).  Accordingly, the question posed in this paper is as follows: \emph{is it possible to leverage ``similarities'' among the $N$ systems to obtain accurate estimates of the system matrices, even if the LS is ill-conditioned? In particular, if one has only a few measurements  for the $i$-th system,  can one use recorded data from the other LTI systems to improve the estimation error?} 

In this direction,~\cite{xin2022identifying} considered estimating  the matrices of a linear system from samples generated by a ``similar'' one;   in~\cite{xin2022identifying}, an LTI system is considered  ``similar'' if its matrices  are perturbed versions of a given matrix. Recently,~\cite{wang2022fedsysid} considered a setup where the matrix norm of the difference between the matrices of LTI systems is small. In this paper, we expand the notion of ``structural similarity'' to account for additional  
properties that the $N$  systems may have in common, and propose a new system identification approach  that leverages and cross-fertilizes core tools investigated in the context of multi-task learning~\cite{evgeniou2004regularized,sener2018multi,zhang2021survey}, statistical learning~\cite{hastie2009elements}, and regularized identification methods~\cite{pillonetto2022regularized,brunton2016discovering}.

We first consider the case where the $N$ LTI systems
model networked systems with a \emph{similar connectivity (sub-)graph}; this implies that the matrices of the LTI systems feature a \emph{common sparsity} (i.e., the system matrices have zeros in a common set of entries). With this model, we formulate the  multi-task system identification task as a \emph{regularized 
LS} problem where we  minimize the LS fit for each of the LTI systems plus a regularization function that enforces a common sparsity pattern~\cite{yuan2006model,huang2010benefit}. By appropriately tuning (typically via cross-validation~\cite{hastie2015statistical}) the weight assigned to the regularization function, one can find a balance between fitting of the recorded data and model complexity. We analyze the estimation error of the proposed multi-task system identification approach, with respect to the true matrices of the LTI systems and with respect to an ``oracle;'' the latter 
represents the best achievable estimation when considering a (group) sparse model, under a given model compatibility condition~\cite{10.5555/2031491}. 

Next, we explain how the proposed  multi-task system identification can be adjusted to account for additional structural similarities. In particular, we provide approaches to deal with cases where the matrix \emph{feature a small heterogeneity} (i.e., the matrix difference is small, as in~\cite{wang2022fedsysid}), and where some of the \emph{matrices of the LTI systems can be expressed as linear combinations} of each other. In this case, we resort to regularization functions that penalize large matrix deviations and functions that are inspired by nuclear norm minimization~\cite{chandrasekaran2009sparse,mardani2015subspace,mohan2010reweighted}.  

We demonstrate the effectiveness of the proposed multi-task system identification method using: (i) synthetic LTI systems that feature structural similarities, and (ii) real data from the Human Connectome Project (HCP), where blood-oxygen-level-dependent (BOLD)  signals are obtained from resting state functional magnetic resonance imaging (fMRI) \cite{srivastava2020models,nozari2020brain}. 
For the latter,  we show that the proposed method requires a significantly smaller number  of fMRI scans to achieve the same error of the LS by simply assuming that the underlying functional or structural connectivity of brain parcellations is similar across   subjects. We also consider the case where only a few fMRI readings are available for one subject, showing the ability to ``transfer information'' from the dynamics of the other subjects.

\section{Preliminaries and System Identification Setup}
\label{sec:prelim}

Consider $N$ linear time-invariant (LTI) systems\footnote{\textbf{Notation}: We  denote by $\mathbb{N}$ and $\mathbb{R}$ the set of natural numbers and the set of real numbers, respectively, and define $[n]=\{1,2,\ldots,n\}$. Given $S \subset [n]$, 
$|S|$ is the cardinality of $S$. We let $^\top$ denote transposition.
For a given column vector $x \in \mathbb{R}^n$, $\|x\|_2$ is the Euclidean norm and $\|x\|_1$ denotes the $\ell_1$ norm; for a matrix $X \in \mathbb{R}^{n \times m}$, $\|X\|_F$ denotes the Frobenious norm and $\|X\|_*$ the nuclear norm. Moreover, $(x)_i$ refers to the entry $i$ of the vector $x$, $(X)_{ij}$  to the entry $(i,j)$ of the matrix $X$, and $\operatorname{vec}(X)$ is a $mn \times 1$ vector stacking the columns of $X$. Given a differentiable function $f: \mathbb{R}^n \rightarrow \mathbb{R}$, $\nabla f(x)$ denotes the gradient of $f$ at $x$ (taken to be a column vector). Given a closed convex set $C \subseteq \mathbb{R}^n$, $\textrm{proj}_{C}:\mathbb{R}^n \to \mathbb{R}^n$ denotes the Euclidean projection of $y$ onto $C$, namely $\textrm{proj}_{C} (y) := \arg \min_{v \in C} \norm{y-v}$.
Given a lower-semicontinuous convex function $g:\mathbb{R}^n \to \mathbb{R}$, the proximal operator is defined as $\textrm{prox}_{\lambda g}(y) := \arg \min_{x \in \mathbb{R}^n} g(x) + \frac{1}{2 \lambda} \|x - y\|_2^2$. }
\begin{align}
\label{eq:system}
x_{i}(t+1) = A_i x_i (t)+ B_i u_i (t) + w_i(t) , \quad x_i(0) \in \mathbb{R}^n,
\end{align}
with $i \in [N]$ the system index and $t \in \mathbb{N}$ the time index, $A_i \in \mathbb{R}^{n \times n}$ and  $B_i \in \mathbb{R}^{n \times p}$, and where $x_i(t) \in \mathbb{R}^{n}$, $u_i(t) \in \mathbb{R}^{p}$, and $w_i(t) \in \mathbb{R}^{n}$ are the state, input and process noise, respectively, of the $i$th system. Assume that, for each system, the input $u_i(t)$ and state $x_i(t)$ can be measured, and $B_i \in \mathbb{R}^{n \times p}$ is known; on the other hand, the system matrix is unknown and the  disturbance $w_i(t) $ cannot be measured\footnote{We note that, while we focus on the estimation of the matrix $A_i$ for notation simplicity and to streamline exposition, the techniques presented in the paper are straightforwardly applicable to the case where both $A_i$ and $B_i$ are estimated from data.}. 

For the $i$-th system, suppose that one has access to one  trajectory $\{x_i(\tau), u_i(\tau)\}_{\tau = 1}^{P_i+1}$, for some $P_i \in \mathbb{N}$, for the state and the inputs. With these measurements, 
the system matrices can be estimated using the following LS criterion:
\begin{align}
\label{eq:ls_i}
 \min_{A_i \in \mathbb{R}^{n \times n}} \mathcal{L}_i(A_i)  ,   
\end{align}
where $\mathcal{L}_i(A_i) := \sum_{\tau = 1}^{P_i} \|x_i(\tau+1) - A_i x_i(\tau) - B_i u_i(\tau)\|_2^2 $; the LS problem~\eqref{eq:ls_i} is solved for each of the $N$ systems independently. The LS estimator~\eqref{eq:ls_i}  has been extensively studied in the literature, especially when the recorded data  render the LS~\eqref{eq:ls_i} well conditioned~\cite{faradonbeh2018finite,simchowitz2018learning,oymak2019non,sarkar2019near,sun2020finite}. 

In this paper, we are interested in cases where the LS problem~\eqref{eq:ls_i} is ill-conditioned for some of the $N$ LTI systems; this may be due to recorded trajectories that are not sufficiently rich, or simply not long enough. In this case, the question we pose  in this paper pertains to whether  it is possible to leverage ``similarities'' among the $N$ systems to obtain accurate estimates of the system matrices, even if the LS is ill-conditioned for some systems. We consider the following structural similarities across the LTI systems: 
\begin{itemize}
    \item[(s1)] The matrices $A_1, \ldots A_N$ have zeros in the same entries; i.e., $(A_1)_{ij} = (A_2)_{ij} = \ldots = (A_N)_{ij} = 0$ for some entries $(i,j)$. 
    \item[(s2)] For any pair $A_i, A_j$, $i,j \in [N]$,  there exists $\epsilon > 0$ such that $\|A_i - A_j\|_F^2 \leq \epsilon$.  
    \item[(s3)]   
  For the subset of systems $i \in \mathcal{C}, \mathcal{C} \subseteq [N]$, there exists $\{\alpha_{i,j} \in \mathbb{R}\}$ such that $A_i = \sum_{j = 1, j\neq i}^N \alpha_{ij} A_j$. 
\end{itemize}
We note that (s1) naturally models LTI systems that  describe the dynamics of networked systems with similar connectivity; as a concrete example, (s1) emerges from a similar functional or structural connectivity of the brain network across different individuals~\cite{srivastava2020models,nozari2020brain}. Similarity, (s2) models the case where the norm of the matrix difference $A_i - A_j$ is small; this is in line with the models considered in~\cite{xin2022identifying,wang2022fedsysid}. 
Finally, (s3) models the case where the matrix $A_i$ of the $i$-th system can be expressed as a linear combination of some of the other matrices $\{A_j\}_{j = 1, j\neq i}^N$; as an example,  this model may  be applicable to traffic flows and mobility-on-demand services (see, e.g.,~\cite{turan2021competition}), where the LTI systems~\eqref{eq:system}  model the evolution of the density of vehicles in given geographical areas over given periods of the day. 

\begin{figure}[t!] 
    \centering
\includegraphics[width=.49\textwidth]{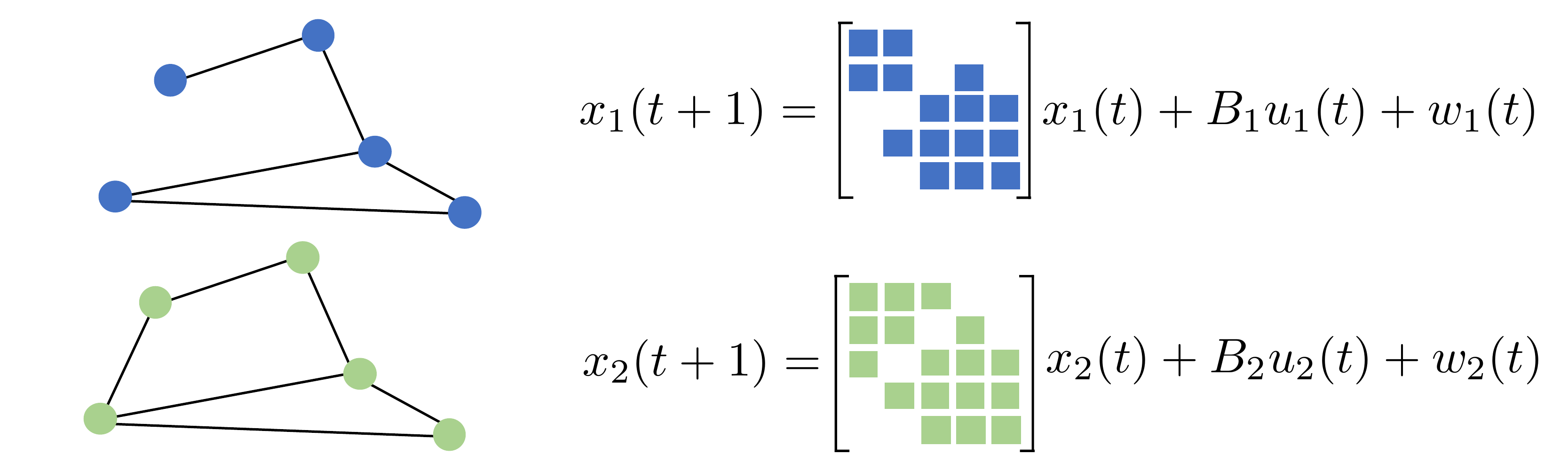}
   \caption{Example of network systems with similar connectivity. the two matrices $A_1$ and $A_2$ feature zeros on a common set of entries.}
    \label{fig:networks}
    \vspace{-0.3cm}
\end{figure}

Given the models (s1)-(s3), we consider estimating the matrices $\{A_i\}_{i \in [N]}$ jointly by solving the following system identification problem:
\begin{align}
\label{eq:mtsysid}
\hspace{-.2cm} \min_{\{A_k\}_{k \in[N]}} \sum_{i = 1}^N  \mathcal{L}_i(A_i)    + \lambda \mathcal{R}(A_1, \ldots, A_N), 
\end{align}
where we recall that $\mathcal{L}_i(A_i)$ is the LS fit for the $i$th system and, in the spirit of regularized regression methods~\cite{hastie2009elements,pillonetto2022regularized,brunton2016discovering}, $(A_1, \ldots, A_N) \mapsto \mathcal{R}(A_1, \ldots, A_N)$ is a lower-semicontinuous  convex function that promotes the prior specified by (s1)--(s3), and $\lambda > 0$ is a tuning parameter. Problem~\eqref{eq:mtsysid} is inspired by multi-task learning methods~\cite{evgeniou2004regularized,sener2018multi,zhang2021survey,crawshaw2020multi}, where learning tasks are performed simultaneously (in our case, the LS fitting) while exploiting commonalities on the parameters that are learned~\cite[Definition~1]{zhang2021survey}. In the ensuing sections, we will explain specific choices for the  regularization function $\mathcal{R}(A_1, \ldots, A_N)$; we start with the case where the $N$ LTI systems feature a common connectivity.

\section{Systems with similar connectivity}

\subsection{Multi-task System Identification}

In this section, we consider the case where the system matrices $A_1, \ldots A_N$ have zeros in a common set  of entries; i.e., $(A_1)_{ij} = (A_2)_{ij} = \ldots = (A_N)_{ij} = 0$ for a subset of indices $i \in [n]$ and $j \in [n]$.  In the statistical learning literature, this structural similarity leads to a setup where the unknowns $\{A_1, \ldots, A_N\}$ feature a \emph{group sparsity}~\cite{yuan2006model,huang2010benefit}. Leveraging the technical approach of~\cite{yuan2006model,huang2010benefit}, we then formulate the multi-task system identification problem for LTI systems  with similar connectivity as follows: 
\begin{align}
\hspace{-.2cm}  \min_{\{A_k\}_{k \in[N]}} &  \sum_{i = 1}^N  \sum_{\tau = 1}^{P_i} \|x_i(\tau+1) - A_i x_i(\tau) - B_i u_i(\tau)\|_2^2 \nonumber \\
& + \lambda \sum_{i = 1}^N \sum_{j = 1}^N \|[(A_1)_{ij}, (A_2)_{ij}, \ldots, (A_N)_{ij}]\|_2 , \label{eq:group_sparsity}
\end{align}
where $\lambda \geq 0$ and the function $\mathcal{R}(A_1, \ldots, A_N) = \sum_{i = 1}^N \sum_{j = 1}^N \|[(A_1)_{ij}, (A_2)_{ij}, \ldots, (A_N)_{ij}]\|_2$ is utilized to enforce group sparsity; that is, the solution of~\eqref{eq:group_sparsity} is such that either  the whole vector $[(A_1)_{ij}, (A_2)_{ij}, \ldots, (A_N)_{ij}]$ is zero or not. The parameter $\lambda$ in~\eqref{eq:mtsysid} strikes a balance between the LS fit (in our case, the LS fit for individual LTI systems) and a number of vectors $[(A_1)_{ij}, (A_2)_{ij}, \ldots, (A_N)_{ij}]$ that are set to zero (as shown shortly).

Problem~\eqref{eq:group_sparsity} is an unconstrained convex program; given the composite cost, we consider a proximal-gradient method (with line search) for solving \eqref{eq:group_sparsity}  (see, e.g., \cite{beck2009gradient,combettes2011proximal}). The proximal-gradient method is tabulated as Algorithm~1. 
\begin{algorithm}[h!]
\caption{Proximal gradient method for systems with similar connectivity}
\begin{algorithmic}
\STATE{Given}:  $\hat A_1^{(0)},\cdots, \hat A_N^{(0)}, \eta^{(0)}$, $\beta \in(0,1)$, and $\lambda$.

\STATE{\textbf{Repeat}:} $m=0,1,2,\ldots$ until convergence 

\quad \textbf{[S1]} $\alpha \leftarrow \eta^{(m)}$.

\STATE{\quad \textbf{[S2]} \emph{Proximal-gradient with line search}:}
 
\quad \quad  \textbf{[S2.1]} $Z_i = \hat A_i^{(m)}-\alpha \nabla\mathcal{L}_i (\hat A_i^{(m)})$, $i \in [N]$

\quad\quad  \textbf{[S2.2]} Let $z_{jk} := [(Z_1)_{jk}, (Z_2)_{jk}, \ldots, (Z_N)_{jk}]$. 

\quad\quad \quad\quad \quad  For all $j,k \in [n]$, compute: 
\begin{align}
\label{eq:glasso}
\hspace{.7cm} y_{jk} = \frac{z_{jk}}{\|z_{jk}\|_2} \max (\|z_{jk}\|_2-\alpha\lambda, 0) .
\end{align}

\quad\quad  \textbf{[S2.3]} Form matrices $\{Y_\ell\}_{\ell\in [N]}$  as $(Y_\ell)_{jk} = (y_{jk})_{\ell}$.

\quad\quad  \textbf{[S2.4]} Break if:  
\small
$ \sum_{i=1}^N \mathcal{L}_i(Y_i)\leq \frac{1}{2\lambda} \|  Y_i- \hat A_i^{(m)}\|_F^2$ 

\quad\quad \quad\quad \quad 
$+\sum_{i=1}^N \left( \mathcal{L}_i(\hat A_i^{(m)}) + \nabla \mathcal{L}_i(\hat A_i^{(m)})^{\top}(Y_i-\hat A_i^{(m)})  \right)
$

\quad \quad   \textbf{[S2.5]} Update $\alpha \leftarrow \beta \alpha$.

\STATE{\quad \textbf{[S3]}} $\eta^{(m+1)} \leftarrow \alpha, \hat A_i^{(m+1)} \leftarrow Y_i$, $i \in [N]$.

\end{algorithmic}
\end{algorithm}

We note that the step [S2.3] is in fact a closed-form expression for the proximal map: 
\begin{align}
\label{eq:prox}
\{Y_i\}_{i \in [N]} =\textrm{prox}_{\alpha\lambda \mathcal{R}}(\{Z_i\}_{i \in [N]}),
\end{align}
and it involves $n^2$ parallel computations as in~\eqref{eq:glasso}.  The convergence behavior of Algorithm~1 can be readily analyzed by leveraging the results in~\cite[Chapter~2]{beck2009gradient}. Moreover, Algorithm~1 can be converted into a ``standard'' proximal-gradient method if the line search is not performed~\cite{parikh2014proximal}. 

From the thresholding operation~\eqref{eq:glasso}, it is clear that increasing $\lambda$ has the effect of forcing a higher number of entries of the system matrices to be zero. Unfortunately, tuning $\lambda$ is not an easy task, and cross-validation procedures are typically utilized to find the value of $\lambda$ such that the estimated matrices yield the lowest error on test data; see, for example~\cite{hastie2015statistical,hastie2009elements}.

\subsection{Analysis}
In this section, we analyze the  performance of the multi-task system identification method. The performance of  regression problems with sparsity-enforcing regularization terms is oftentimes compared against an ``oracle'' that represents the \emph{best achievable estimation} when considering a (group) sparse model, under a given model compatibility condition~\cite{10.5555/2031491}. We will provide error bounds with respect to both the oracle and the true system matrices.

To simplify the notation, we outline the results for the case where $P_i=P$ for all $i \in [N]$ (though, similar results hold when the $\{P_i\}$ are different).  Let $A^S \in \mathbb{R}^{n \times n}$  be a matrix formed by selecting the entries of $A \in \mathbb{R}^{n \times n}$ indexed by $S \subset [n]\times [n]$ and setting zero to the other entries.  For $N$ matrices $A_i \in \mathbb{R}^{n \times n}$, $i \in [N]$, define for brevity $\|\{A_i\}_{i\in[N]}\|_{2,1}=  \sum_{i = 1}^n \sum_{j = 1}^n \|[(A_1)_{ij}, (A_2)_{ij}, \ldots, (A_N)_{ij}]\|_2$. With this notation in place, we first state the main assumptions and outline the compatibility condition associated with the group sparse model~\cite[Chapter~8]{10.5555/2031491}. 

\vspace{.1cm}

\begin{assumption}
\label{as:noise}
The disturbances $\{w_i(k)\}$ are i.i.d. Gaussian random variables $\mathscr{N}(0,\sigma^2)$. \hfill $\Box$ 
\end{assumption}

\vspace{.1cm}

\begin{definition}[\cite{10.5555/2031491}]
Let $S \subset [n]\times [n]$ and $S^c := ([n]\times [n]) \setminus S$. 
 We say that the \emph{compatibility condition} holds for the index set $S$ if for any $\{A_i \in \mathbb{R}^{n \times n}\}_{i \in [N]}$ with $\|\{A_i^{S^c}\}_{i\in[N]}\|_{2,1} \leq 3\|\{A_i^{S}\}_{i\in[N]}\|_{2,1}$, it holds that
\begin{align}
\label{eq:comptibility}
\|\{A_i^{S}\}_{i\in[N]}\|_{2,1}^2 \leq \frac{ |S| \sum_{i=1}^N\sum_{\tau = 1}^{P} \|A_i x_i(\tau) \|_2^2   }{P\phi(S)}
\end{align}
for some constant $\phi(S)>0$. Moreover, we define as  $\mathcal{S}$ the collection of sets $S$ for which the  compatibility condition holds.  \hfill $\Box$  
\end{definition}
\vspace{.1cm}

We note that, when $S$ coincides with the support of the matrices $\{A_i\}_{i \in [n]}$, this  technical condition provides bounds on the values of the non-zero entries of the matrices.  

Hereafter, we let $\{A_i^{MT}\}_{i \in [N]}$ be an optimal solution of the system identification problem~\eqref{eq:group_sparsity}, and we denote as $\{A_i^{\star}\}_{i \in [N]}$ the \emph{true} matrices of the LTI systems in~\eqref{eq:system}. 
The following theorem provides an error bound for the proposed multi-task system identification~\eqref{eq:group_sparsity} relative to the (true) system matrices $\{A_i^{\star}\}_{i \in [N]}$. 

\vspace{.1cm}

\begin{theorem}
\label{thm:bounds_true} 
Let $\{A_i^{\star}\}_{i \in [N]}$ be the true matrices of the LTI systems and define $S^\star=\{(i,j)\mid [(A_1^\star)_{ij}, (A_2^\star)_{ij}, \ldots, (A_N^\star)_{ij}]\neq 0 \}$. Let Assumption~\ref{as:noise} hold and suppose that $S^\star \in \mathcal{S}$. 
Define
\begin{align}
\label{eq:lambda0}
\lambda_0:=\frac{2 \sqrt{M}}{nP}\left(1+\sqrt{\frac{4 \gamma+8 \log n}{N}}+\frac{4 \gamma+8 \log n}{N}\right)^{\frac{1}{2}} 
\end{align}
for some $\gamma>0$, where $M = \sigma^2 \max\limits_{1\leq i\leq N,1\leq j\leq n} \sum_{k=1}^P [(x_i^{(j)})_k]^2$. If $\lambda\geq 4nNP\lambda_0$, then,  the following bound holds
\begin{align}
&\sum_{i = 1}^N  \sum_{\tau = 1}^{P} \|(A_i^\star- A_i^{MT}) x_i(\tau) \|_2^2  \leq \frac{24 \lambda^2 |S^\star| }{PN\phi(S^\star)} \label{eq:bound_astar}
\end{align}
with probability at least $1-\mathrm{e}^{-\gamma}$. \hfill $\Box$
 \end{theorem}

\vspace{.1cm}

From Theorem \ref{thm:bounds_true}, it can be seen that the average error $E(N,P) := \frac{1}{PN}\sum_{i=1}^N\sum_{i = 1}^N  \sum_{\tau = 1}^{P} \|(A_i^\star- A_i^{MT}) x_i(\tau) \|_2^2 $ is of the order 
$$E(N,P) \approx O\left(\frac{|S^\star|}{\phi(S^\star)} P^{-1} (1+ N^{-\frac{1}{2}}+ N^{-1})\right),$$
when taking $\lambda=O(\sqrt{P}\sqrt{N+\sqrt{N}+1})$. Interestingly, by increasing the number of LTI systems in  the multi-task system identification (i.e. $N$ increases), the average  error $E(N,P)$ decreases; however, when $N \rightarrow \infty$, the error does not tend to $0$. This can be understood as a plateau in the ability to ``transfer information'' between systems. 

One possible shortcoming of Theorem~\ref{thm:bounds_true} is that the set $S^\star$ describing the (common) support of the matrices $\{A_i^{\star}\}_{i \in [N]}$ is assumed to satisfy the compatibility condition.  In the following, we offer an additional error bound for cases where  $\{A_i^{\star}\}_{i \in [N]}$ is not guaranteed to satisfy the compatibility condition; the error bound leverages the notion of oracle~\cite{10.5555/2031491}. 

\vspace{.1cm}

\begin{theorem}
\label{thm:bounds_oracle} 
Consider the \emph{oracle} $\{A_i^\dagger\}_{i \in [N]}$ defined as:
\begin{align}
\{A_i^\dagger\}_{i \in [N]} \in \arg\min _{ \{A_i\}: S_{\{A_i\}} \in \mathcal{S}}&\left\{\sum_{i=1}^N\sum_{\tau = 1}^{P} \|(A_i^\star- A_i) x_i(\tau) \|_2^2\right. \nonumber \\
& \hspace{.6cm} \left.+\frac{4  \lambda^2 |S_{\{A_i\}}|}{PN\phi\left(S_{\{A_i\}}\right)}\right\},  \hspace{-.2cm}
\end{align}  
where $S_{\{A_i\}} =\{(i,j)\mid [(A_1)_{ij}, (A_2)_{ij}, \ldots, (A_N)_{ij}]\neq 0 \} $, and let Assumption~\ref{as:noise} hold. Set $\lambda_0$ and $\lambda$ as in Theorem~\ref{thm:bounds_true}. Then, the following bound holds with  probability at least $1-\mathrm{e}^{-\gamma}$, for a given $\gamma > 0$: 
\begin{align}
&\sum_{i = 1}^N  \sum_{\tau = 1}^{P} \|(A_i^\star- A_i^{MT}) x_i(\tau) \|_2^2  \nonumber \\
& \hspace{0.5cm}\leq  6\sum_{i = 1}^N  \sum_{\tau = 1}^{P} \|(A_i^\star- A_i^\dagger) x_i(\tau) \|_2^2  +\frac{24 \lambda^2 |S^\dagger| }{PN\phi(S^\dagger)}, \label{eq:bound_oracle}
\end{align}
where $S^\dagger =\{(i,j)\mid [(A_1^\dagger)_{ij}, (A_2^\dagger)_{ij}, \ldots, (A_N^\dagger)_{ij}]\neq 0 \} $. 
\hfill $\Box$
 \end{theorem}

\vspace{.1cm}

Theorem~\ref{thm:bounds_oracle} asserts that the  error  incurred by the multi-task system identification~\eqref{eq:group_sparsity} is bounded by the estimation error associated with the oracle plus an additional term modeling the error between the estimated matrices and the oracle. Here, the oracle represents the best achievable estimation when considering matrices with support index set that satisfies the compatibility condition.

\section{Handling Systems with Other Similarities}

In this section, we explain how the proposed multi-task system identification method can be adapted to LTI systems that feature additional similarities, as previously explained in Section~\ref{sec:prelim}.   

\subsection{Small Heterogeneity}

Consider the case where, for any pair $A_i, A_j$, $i,j \in [N]$,  there exists $\epsilon > 0$ such $\|A_i - A_j\|_F^2 \leq \epsilon$. This case is referred to as ``small heterogeneity'' in~\cite{wang2022fedsysid}. With this prior, the multi-task system identification  problem can be formulated as 
\begin{align}
 \min_{\{A_k\}_{k \in [N]}} \sum_{i = 1}^N \mathcal{L}_i(A_i) + \lambda \sum_{i = 1}^N \sum_{j = i}^N \|A_i - A_j\|_F^2, \label{eq:deviations}
\end{align}
where we recall that $\mathcal{L}_i(A_i)$ is the LS fit for the $i$th system and where the regularization term penalizes large deviations between the estimated matrices~\cite{hastie2009elements}.

Problem~\eqref{eq:deviations} is convex and can be solved in closed form. However, to avoid computationally-heavy matrix inversions (especially when the dimension $b$ is large and several systems are considered in the system identification process), Algorithm~1 can be utilized to solve~\eqref{eq:deviations} by replacing  [S2.2] with the following $n^2$ parallel computations:  
 \begin{align}
 \label{eq:diff}
 y_{ij}= \frac{z_{ij}+2\alpha\lambda s_{ij}[1,1,\cdots,1]}{2\alpha\lambda N+1},  \quad i,j \in [N] ,
 \end{align}
 where $s_{ij}=\sum_{\ell=1}^N (Z_\ell)_{ij}$. 
 
 From the update~\eqref{eq:diff}, it can be seen that $y_{ij} \rightarrow (s_{ij}/N)[1,1,\cdots,1]$ as $
 \lambda \rightarrow \infty$, thus setting all the system matrices to be the same. On the other hand, by setting $\lambda = 0$ one recovers the LS method. Even in this case,  cross-validation procedures can be  utilized to find the value of $\lambda$ such that the estimated matrices yield the lowest error on test data~\cite{hastie2015statistical}. 

The estimation performance of~\eqref{eq:deviations} can be analyzed by deriving bounds between an optimal solution of ~\eqref{eq:deviations} and the one of the LS method. Since the bound is straightforward, and because of space limitations, we omit this result from the paper.

\subsection{Linear Combinations}

Lastly, we comment on an additional ``similarity'' where the system matrices are (approximately) linearly dependent. Precisely, we consider a scenario where for a subset of systems, there exists coefficients $\{\alpha_{i,j} \in \mathbb{R}\}$ with $\alpha_{i,j} \neq 0$ for some $j \in [N]$ such that $A_i \approx \sum_{j = 1, j\neq i}^N \alpha_{ij} A_j$. 

To formalize the setup, suppose that 
$q \ll N$ of the matrices $\{A_i\}_{i \in [N]}$ are such that the remaining $N-q$ can be represented as a linear combination of these $q$ matrices. Consider then building the  $n^2 \times N$ matrix $[\operatorname{vec}(A_1),\operatorname{vec}(A_2),\ldots,\operatorname{vec}(A_N)]$; it follows that this matrix 
has rank $q \ll N$. Based on this observation, we propose to formulate a multi-task system identification  problem for this case as 
\begin{align*}
 \min_{\{A_k\}_{k \in [N]}} &\sum_{i = 1}^N  \mathcal{L}_i(A_i) \\
 &+ \lambda \left\|[\operatorname{vec}(A_1),\operatorname{vec}(A_2),\ldots,\operatorname{vec}(A_N)] \right\|_* ,\label{eq:combinations}
\end{align*}
where the regularization function promotes sparsity in the singular values of the matrix $[\operatorname{vec}(A_1),\operatorname{vec}(A_2),\ldots,\operatorname{vec}(A_N)]$
 (see, e.g.,~\cite{chandrasekaran2009sparse,mardani2015subspace}). 

 Problem~\eqref{eq:combinations} is convex and with a composite cost where the  regularization function is not differentiable~\cite{chandrasekaran2009sparse,mardani2015subspace}. Still, the proximal-gradient algorithm tabulated as Algorithm~1 can be modified to solve~\eqref{eq:combinations}. In particular,
 one can replace the step [S2.2] is Algorithm~1 with $\textrm{prox}_{\alpha\lambda \left\|[\operatorname{vec}(A_1),\operatorname{vec}(A_2),\ldots,\operatorname{vec}(A_N)] \right\|_*} (\{Z_i\}_{i \in [N]})$; this proximal map affords a closed-form solution given by: 
\begin{align}
\bar Y = U \textrm{diag} ( \{\max\{\sigma_i-\alpha\lambda,0 \} \}) V^*,
\end{align}
where  the singular value decomposition of $[\operatorname{vec}(A_1),\operatorname{vec}(A_2), \ldots,\operatorname{vec}(A_N)]$ is $U \textrm{diag} (\{\sigma_i \}) V$.
The matrices $\{Y_\ell\}_{\ell\in [N]}$ are then extracted from the columns of $\bar Y$. From the computation of $\bar Y$, it can be seen that higher values of $\lambda$ lead to a higher number of  singular values of $\bar Y $ that are set to zero. 

While the effectiveness of nuclear norm minimization approaches has been verified numerically in the literature, identifiability results and analytical bounds for the estimation error are available only under given assumptions on the regressors~\cite{chandrasekaran2009sparse,mardani2015subspace} that may not be applicable to~\eqref{eq:combinations}; see also~\cite{mohan2010reweighted}. Deriving analytical error bounds for~\eqref{eq:combinations} is subject of our ongoing investigations.       

\vspace{.1cm}

\begin{remark}
The proposed multi-task system identification problems can be extended to cases where the system matrices $\{A_i\}_{i \in [N]}$ are similar according  to more than one of the priors (s1)--(s2). For example, if the matrices have a common sparsity pattern and the differences in the non-zero entries are small, one can utilize the composite regularization function $\lambda_1 \sum_{i = 1}^N \sum_{j = i}^N \|A_i - A_j\|_F^2$ $+ \lambda_2 \sum_{i = 1}^N \sum_{j = 1}^N \|[(A_1)_{ij}, (A_2)_{ij}, \ldots, (A_N)_{ij}]^\top\|_2$, where $\lambda_1, \lambda_2 \geq 0$ are tuning parameters. \hfill $\Box$
\end{remark}


\section{Numerical Experiments}

\subsection{Experiments on brain networks}

We test the proposed method for the problem of estimating the dynamics of brain networks, using data corresponding to the resting state fMRI  from the Human Connectome Project (HCP)\footnote{Data available at https://wiki.humanconnectome.org/} \cite{nozari2020brain,srivastava2020models,gu2015controllability}. Here, $x_i(t)$ is a $116$-dimensional blood-oxygen-level-dependent (BOLD) time series for $116$ parcellations of the brain of the $i$-th  subject. Our goal here is to estimate $N=5$ dynamical systems of the form $x_{i}(t+1) = A_i x_i (t) + w_i(t)$, that model the evolution of BOLD signal when the individual is in a resting state, with $w_i(t)$ capturing process noise (the model does not contain external inputs $u_i$ due to the resting state condition).

Since the matrices $\{A_i\}_{i \in [5]}$  are unknown, we consider the following error for each system:
$$
\mathcal{E}(A) : =\frac{1}{n} \sum_{k=1}^{n} \frac{\sum_{i=1}^{p} (x_i(k) -[A x_i](k) )^2  }{ \sum_{i=1}^{p} (x_i(k) -\bar{x}(k) )^2   },
$$
where  $n$ is the length of the testing vector, $p$ is the number of testing data and $\bar{x}(k) : =\frac{1}{p} \sum_{i=1}^p x_i(k)$. Note that $1-\mathcal{E}(A) $ is precisely the average $R^2$ indicator of~\cite{nozari2020brain}. 

We consider three different methods: (i) the LS estimator~\eqref{eq:ls_i}, which is utilized per individual; (ii) the  Least Absolute Shrinkage and Selection Operator (LASSO), which is again utilized per individual as proposed in~\cite{nozari2020brain}; and, (iii) the proposed method~\eqref{eq:mtsysid} with the 
group-sparsity regularization function~\eqref{eq:group_sparsity}. The rationale behind the group-sparsity is that the brain dynamics should exhibit the same effective connectivity between parcellations, though the remaining entries acknowledge the diversity in intensities of the interactions across individuals. We note that the effectiveness of the LS and LASSO has been experimentally validated in~\cite{nozari2020brain}, where their estimation accuracy has been compared with several identification methods. Moreover, we performed a cross-validation procedure to optimize the performance of the LASSO.   

\begin{figure}[h!]
  \centering 
  \begin{subfigure}[]{\includegraphics[width=0.42\textwidth]{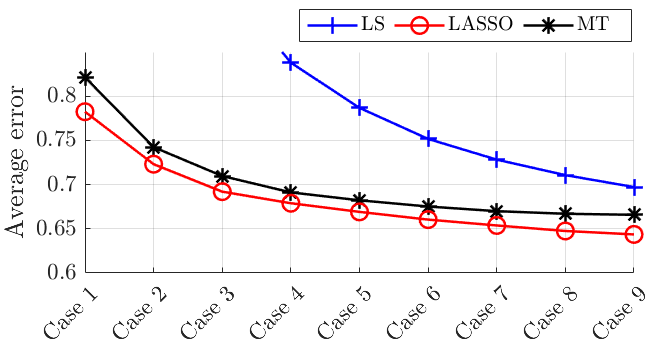}} \end{subfigure}
  \begin{subfigure}[]{\includegraphics[width=0.42\textwidth]{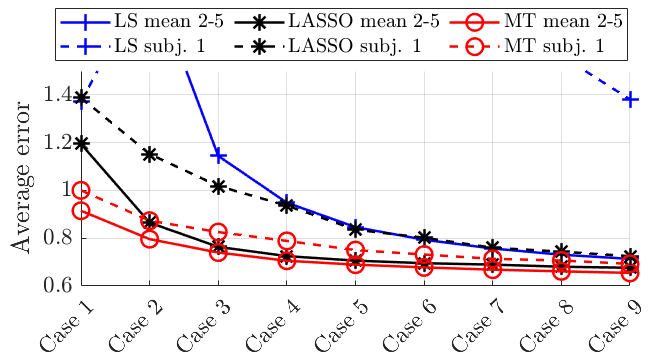}}
  \end{subfigure}
  \vspace{-.2cm}
  \caption{(a) Mean error of LS, LASSO and multi-task (MT) system identification; ``Case $k$'' means that $100 k$ training data points are available for each subject ($k=1,2,\cdots,9$). (b) Mean error for subjects 2-5 and error for subject 1. ``Case $k$'' means that  $25 k$ fMRI scans are used for subjects 1 (dashed line) while $100 k$ (solid line) scans are used for subjects 2-5.}
  \label{fig:mean_error}
\end{figure}

\begin{figure*}[!ht]
  \centering 
  \begin{subfigure}[]{\includegraphics[width=1\textwidth]{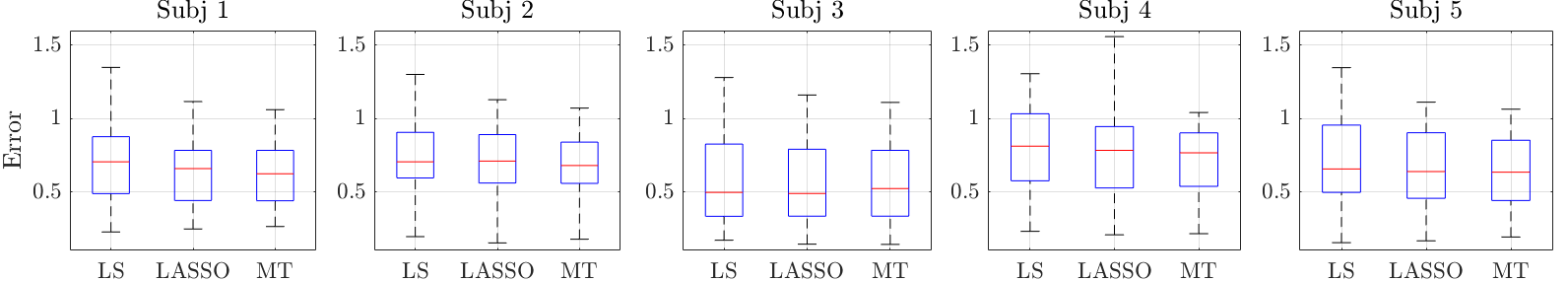}} \end{subfigure}
  \begin{subfigure}[]{\includegraphics[width=1\textwidth]{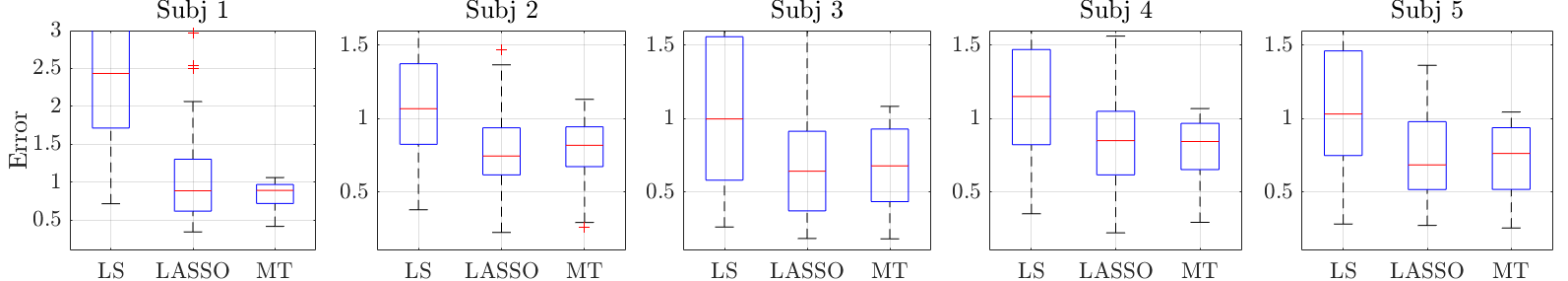}}
  \end{subfigure}
  \begin{subfigure}[]{\includegraphics[width=1\textwidth]{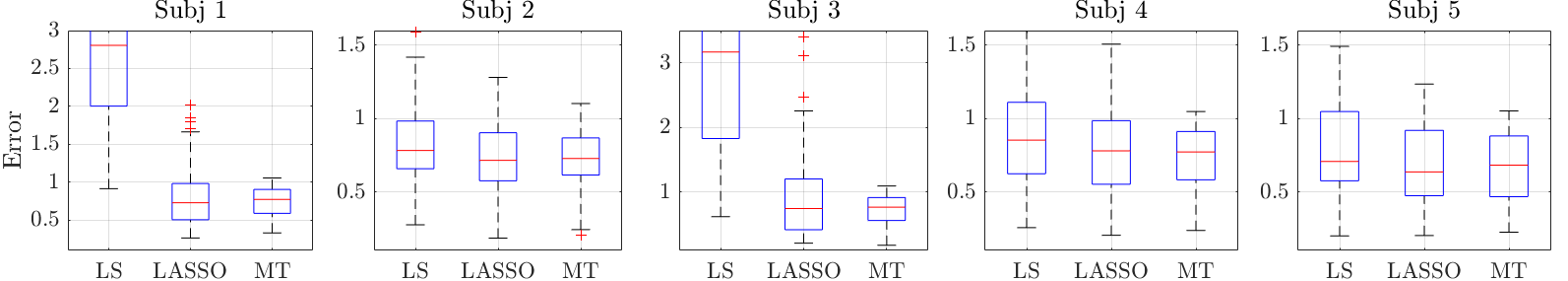}}
  \end{subfigure}
  \vspace{-0.5cm}
  \caption{Comparison between LS, LASSO and multi-task (MT) system identification (a) Case 1: $900$ training data points for each subject. (b) Case 2: For subject 1, $75$ training data points, and $300$ for subjects 2-5. (c) Case 3: For subject 1 and 3, $150$  training data points, and $600$ for subjects 2, 4, and 5. In the box plots, the red center line, box limits, and whiskers represent the median, upper and lower quartiles, and the smallest and largest samples, respectively. Red crosses indicate outliers.}
  \label{fig:comp_methods}
\end{figure*}
\begin{figure*}[h!] 
    \centering
\includegraphics[scale=0.45]{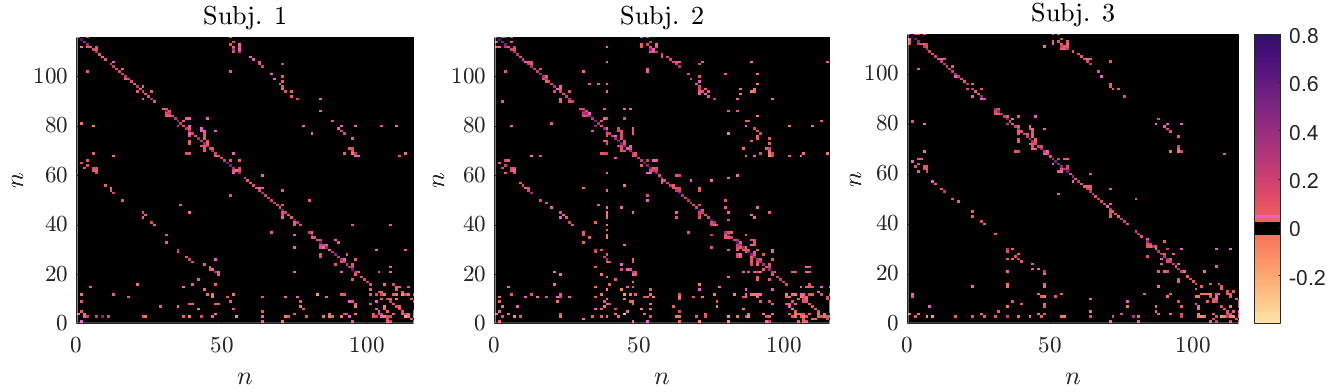}
   \caption{Estimated matrix $\hat A_i$ for Case 3, individuals 1, 2 and 3, for $n = 116$ brain parcellations.}
    \label{fig:heatmap1}
    \vspace{-0.3cm}
\end{figure*}
In Figure \ref{fig:mean_error}, we compare the LS, LASSO and our approach (which is labeled as ``MT'') in two cases: (a) the same amount of training data is utilized for the  five subjects; and, (b) for subject 1, we utilize only 25\% of the training data points  with respect to the other subjects 2-5. We use $100$ test points. In Figure \ref{fig:mean_error}(a) we plot the mean error across the subjects 1-5; in Figure \ref{fig:mean_error}(b) we plot the mean error across the subjects 2-5 and the error for subject 1, for which fewer fMRI readings are available. The proposed method outperforms the LS and the LASSO, on par with the number of fMRI scans in both cases. The merits of the proposed method are particularly evident in Figure~\ref{fig:mean_error}(b), where the proposed method significantly outperforms the LASSO for the subject 1; on the other hand, the LS is ill-conditioned and does not return meaningful estimates. This  shows the ability to leverage information and data (in this case, fMRI readings) from the dynamics of subjects 2-5 to assist the estimation of the dynamics in subject 1. 

To provide additional comparisons other than the mean error, Figure \ref{fig:comp_methods} shows the box plots for the LS, the LASSO, and the proposed approach in three different scenarios. In particular, Figure \ref{fig:comp_methods}(a) shows that proposed multi-task identification method can achieve a smaller or comparable error (on average) than LS or LASSO when trajectories of 900 time steps are used for each subject (and these training trajectories are sufficiently rich). Figure \ref{fig:comp_methods}(b) considers the case where $75$ training data points are available for subject 1 and $300$ for subjects 2-5. Here, the LS does not perform well due to ill-conditioning. The performance of the LASSO is comparable with the one of the proposed method in terms of median; however, the proposed method shows smaller upper and lower quartiles. Moreover, Figure \ref{fig:comp_methods}(c) considers the case where fewer fMRI readings are available for subjects 1 and 3; the proposed method performs better than the LASSO in terms of quartiles and has a significantly less error deviation across the parcellations. 

Finally, a representative example of the estimated matrices $\hat A_i$ for the subjects 1-3 is provided in Figure \ref{fig:heatmap1}. The estimated matrices are the ones obtained in the case considered in Figure \ref{fig:comp_methods}(c), where subjects 1 and 3 have fewer training points. It is possible to notice that the three matrices have zeros in many common entries. Based on this result, we will explore additional regularization methods that will combine group sparsity with (entry-wise) sparsity.

\subsection{Experiments on synthetic data}
We provide additional results on synthetic data. We consider 10 systems as in \eqref{eq:system}, where $\{A_i\}_{i \in [10]} \in \mathbb{R}^{50 \times 50}$, $\{B_i\}_{i \in [10]} \in\mathbb{R}^{50 \times 4}$, $u_i(t)$ is the vector of all ones in $\mathbb{R}^{4}$, i.e. $u_i(t)$ is constant vector and $w_i(t)\sim \mathcal{N}(0,0.1^2)$. We consider two different cases: common sparsity and linear combinations. We compare the LS estimator~\eqref{eq:ls_i} and the proposed method~\eqref{eq:mtsysid} with the group-sparsity regularization~\eqref{eq:group_sparsity} and nuclear norm regularization~\eqref{eq:combinations}.

\begin{figure}[h!]
  \centering 
  \begin{subfigure}[]{\includegraphics[width=0.4\textwidth]{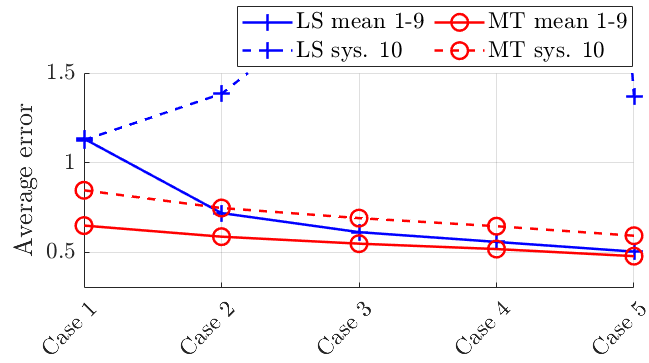}} \end{subfigure}
  \begin{subfigure}[]{\includegraphics[width=0.4\textwidth]{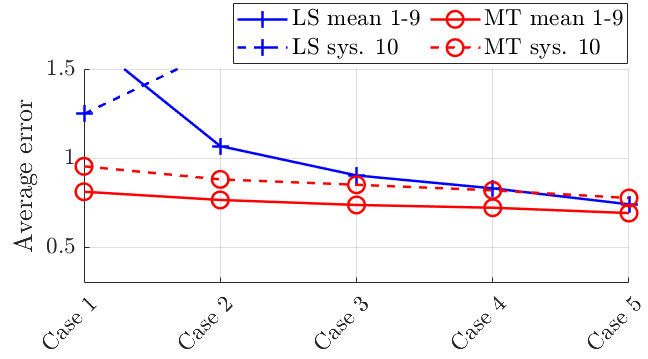}}
  \end{subfigure}
  \vspace{-.3cm}
  \caption{Mean error curve to compare LS and multi-task (MT) system identification methods. (a) Linear combinations. (b) Common sparsity. ``Case $k$'' means that  $10 k+10$ samples of the trajectory are used for system 10 (dash line) while $40k+40 k$ (solid line) are used for system 1-9, $k=1,2,3,4$. In ``Case $5$'', $75$ data points are used for system 10 (dashed line) while $300 k$ (solid line) are used for system 1-9. 
}
  \label{fig:mean_error_sd}
\end{figure}

Figure~\ref{fig:mean_error_sd} compares the LS and our approach in two cases: (a) all the 10 systems can be represented by a linear combination of 3 systems and only 25\% of the training data points are accessible for the tenth system with respect to the other systems 1-9; (b) all the 10 systems have the same sparsity pattern and only 25\% of the training data points are accessible for the tenth system with respect to the other systems 1-9. The testing is on $60$ data points. In Figure~\ref{fig:mean_error_sd}(a), we plot the mean error across systems 1-9 as well as the error for system 10. The proposed method outperforms the LS approach in both the mean error and the error for system 10, especially in the case of only a small number of data available. In Figure~\ref{fig:mean_error_sd}(b), we can observe similar results. 

\bibliographystyle{IEEEtran}
\bibliography{references.bib}

\section*{Acknowledgements}
The authors would like to thank Dr. Erfan Nozari (University of California at Riverside) for the assistance with the data used in the simulations, and Dr. Stephen Becker (University of Colorado Boulder) 
 for the discussion on the group Lasso.

\appendix

\emph{Proofs of Theorems~\ref{thm:bounds_true} and~\ref{thm:bounds_oracle}}. We provide a sketch of the proofs of Theorems~\ref{thm:bounds_true} and~\ref{thm:bounds_oracle}. To this end, we introduce some additional notation. For the $i$th system, collect the recorded trajectories in the matrix $X_i :=[x_i(1),...,x_i(P_i)]$; then, the LS fit $\mathcal{L}_i(A_i)$ for the $i$th system can be equivalently expressed  as 
$$ \mathcal{L}_i(A_i) =  
\|\Tilde{\by}_i - \Tilde{X}_i \Tilde{\ba}_i  \|_2^2 ,$$ where $\Tilde{\by}_i := \operatorname{vec}([x_i(2)-B_iu_i(1),...,x_i(P_i+1)-B_iu_i(P_i)])$, $\Tilde{\ba}_i=\operatorname{vec}(A_i)$ and $\Tilde{X}_i=X_i^{\top} \otimes I_n$. Moreover, let
$\mathbf{\Tilde{A}}:=[\Tilde{\ba}_1,...,\Tilde{\ba}_N]$ collect all the vectorized system matrices $\{A_i\}_{i \in [N]}$, 
$\bx_{i}^{(j)}$ the $j$th column of $\Tilde{X}_i$, and denote as 
$\|\mathbf{\Tilde{A}} \|_{2,1}$ the sum of the $l_2$-norm of each row of $\mathbf{\Tilde{A}}$.   
Finally, recall that for any matrix $\mathbf{\Tilde{A}} \in \mathbb{R}^{n^2 \times N}$,  $\mathbf{\Tilde{A}}_S$ is a matrix formed by selecting the rows of $\mathbf{\Tilde{A}}$ indexed by $S$ and setting to zero the other rows.

With this notation in place, and recalling that $\{A_i^{MT}\}_{i \in [N]}$ denotes an optimal solution of~\eqref{eq:group_sparsity}, we have the following result.

\vspace{.1cm}

\begin{lemma}
\label{thm:bounds} 
Let Assumption~\ref{as:noise} hold.  Let $\lambda_0$ be defined as in Theorem~\ref{thm:bounds_true}, and take $\lambda\geq 4nNP\lambda_0$.  Then, for any $\mathbf{\Tilde{A}} \in \mathbb{R}^{n^2 \times N}$ and any $S \in \mathcal{S} $ satisfying the compatibility condition, the following bound holds with probability at least  $1-\mathrm{e}^{-\gamma}$:
\begin{align}
&\sum_{i=1}^N\|\Tilde{X}_i \Tilde{\ba}_i^{\star} - \Tilde{X}_i \Tilde{\ba}_i^{MT}\|^2_2+\frac{\lambda}{\sqrt{N}}\| \mathbf{\Tilde{A}}^{MT}-\mathbf{\Tilde{A}}_S\|_{2,1}  \nonumber \\
& \hspace{1.5cm} \leq 6\sum_{i=1}^N\left\|\Tilde{X}_i \Tilde{\ba}_i^{\star} - \Tilde{X}_i \Tilde{\ba}_{S,i}\right\|^2_2+\frac{24 \lambda^2 |S| }{PN\phi^2(S)} \label{eq:bound}
\end{align}
for any given $\gamma > 0$. \hfill $\Box$
 \end{lemma}

\vspace{.1cm}

The proof of Lemma~\ref{thm:bounds} is omitted due to space limitations. Concisely, the proof of  Lemma~\ref{thm:bounds} leverages some arguments from Chapters~6 and~8 in~\cite{10.5555/2031491}. 
First, we establish basic inequalities for our problem similarly to~\cite[Chapter~6]{10.5555/2031491}, and bound the empirical process by ~\cite[Lemma.~8.5]{10.5555/2031491}. Then, we derive an inequality similar to ~\cite[Lemma.~6.3]{10.5555/2031491} for our group sparse model. Finally, combining these inequalities, we prove the inequalities in Lemma 1 (similar to ~\cite[Thm.~6.2]{10.5555/2031491}).  The full proof will be made available on an extended version online.

Based on the result of Lemma~\ref{thm:bounds}, the bound in Theorem~\ref{thm:bounds_true} can be shown by setting $\mathbf{\Tilde{A}}_S = \mathbf{\Tilde{A}}^\star$ in~\eqref{eq:bound}, thus obtaining
\begin{align}
&\sum_{i=1}^N\|\Tilde{X}_i \Tilde{\ba}_i^{\star} - \Tilde{X}_i \Tilde{\ba}_i^{MT}\|^2_2+\frac{\lambda}{\sqrt{N}}\| \mathbf{\Tilde{A}}^{MT}-\mathbf{\Tilde{A}}^\star\|_{2,1}  \nonumber \\
& \hspace{5.5cm} \leq \frac{24 \lambda^2 |S^\star| }{PN\phi^2(S^\star)} \label{eq:bound2}
\end{align}
and noticing that $\sum_{i=1}^N\|\Tilde{X}_i \Tilde{\ba}_i^{\star} - \Tilde{X}_i \Tilde{\ba}_i^{MT}\|^2_2 = \sum_{i = 1}^N  \sum_{\tau = 1}^{P} \|(A_i^\star- A_i^{MT}) x_i(\tau) \|_2^2$. 
On the other hand, the bound in Theorem~\ref{thm:bounds_oracle} can be shown by setting $\mathbf{\Tilde{A}}_S = \mathbf{\Tilde{A}}^\dagger$ in~\eqref{eq:bound}, where we recall that $\mathbf{\Tilde{A}}^\dagger$ is the oracle solution.  

\end{document}